\documentclass[letterpaper, 10 pt, conference]{ieeeconf} 
\usepackage[utf8]{inputenc}
\usepackage{verbatim}
\usepackage{float}
\usepackage{amsmath}
\usepackage{amssymb}
\usepackage{graphicx}
\usepackage{esint}

\usepackage{amsmath}
\usepackage{amssymb}
\usepackage{float}
\usepackage{graphicx}

\makeatletter
\floatstyle{ruled}
\newfloat{algorithm}{tbp}{loa}
\providecommand{\algorithmname}{Algorithm}
\floatname{algorithm}{\protect\algorithmname}
\usepackage{algorithm,algpseudocode}
\usepackage{calc}
\usepackage{colortbl}
\newcolumntype{M}[1]
 {>{\centering\hspace{0pt}}m{#1}}
\newcolumntype{S}[2]
 {>{\centering\hspace{0pt}}m{(#1+(2\tabcolsep+\arrayrulewidth)*(1-#2))/#2}}
\newcolumntype{K}[1]
 {>{\columncolor{#1}\hspace{0pt}}c}
\@ifundefined{showcaptionsetup}{}{%
 \PassOptionsToPackage{caption=false}{subfig}}
\usepackage{subfig}
\makeatother

\IEEEoverridecommandlockouts 
\begin{document}

\title{\LARGE \bf Adaptive Critic Based Optimal Kinematic Control for a Robot Manipulator}

\author{Aiswarya Menon$^{1}$, Ravi Prakash$^{2}$, Laxmidhar Behera$^{3}$,~\IEEEmembership{Senior Member,~IEEE}% <-this % stops 
\thanks{All authors are with the Department of Electrical Engineering,
Indian Institute of Technology, Kanpur, India-208016. Email id : $^{1}$aiswarya@iitk.ac.in, $^{2}$ravipr@iitk.ac.in,$^{3}$lbehera@iitk.ac.in%
}}
\maketitle

\begin{abstract}

This paper is concerned with the optimal kinematic control of a robot manipulator where the robot end effector position follows a task space trajectory. The joints are actuated with the desired velocity profile to achieve this task. This problem has been solved using a single network adaptive critic (SNAC) by expressing the forward kinematics as input affine system. Usually in SNAC, the critic weights are updated using back propagation algorithm while little attention is given to convergence to the optimal cost. In this paper, we propose a critic weight update law that ensures convergence to the desired optimal cost while guaranteeing the stability of the closed loop kinematic control. In kinematic control, the robot is required to reach a specific target position. This has been solved as an optimal regulation problem in the context of SNAC  based kinematic control. When the robot is required to follow a time varying task space trajectory, then the kinematic control has been framed as an optimal tracking problem. For tracking, an augmented system consisting of tracking error and reference trajectory is constructed and the optimal control policy is derived using SNAC framework. The stability and performance of the system under the proposed novel weight tuning law is guaranteed using Lyapunov approach. The proposed kinematic control scheme has been validated in simulations and experimentally executed using  a real six degrees of freedom (DOF) Universal Robot (UR) 10 manipulator.
\end{abstract}
\vspace{-2mm}

\section{Introduction}

Modern robotic systems are becoming complex as the degrees of freedom are increasing to address the complex application scenario such as agriculture, health-care and ware-house automation. A mobile manipulator with Shadow hand mounted on it has 28 degrees of freedom. The inverse kinematic solution to such systems can not be solved analytically. Thus neural and fuzzy neural network based schemes have become popular to design kinematic control\cite{Swagat10,Prem10,li2017kinematic,indrazno14}. These approaches can not learn the inverse kinematics while optimizing a global cost function. This is very important when one deals with redundant manipulators. One of the approaches to design optimal kinematic control is to use the Hamilton-Jacobi-Bellman (HJB) formulation \cite{lewis2012optimal, keerthi1985existence}. The analytical solution of the HJB equation is still a
major challenge and as such these solutions are obtained off-line. The real time approximate solutions are known as approximate dynamic programming and are presented in  \cite{bertsekas2005dynamic, chen2008generalized, si2004handbook}. In these approaches, the framework uses two neural networks - one for action and the other critic. Action network learns to actuate an optimal policy while critic network evaluates the cost function through learning.

The single network adaptive critic (SNAC) is introduced  \cite{padhi2006single} where critic
is updated using back-propagation. 
\begin{figure}[!h]
 \centering
 \captionsetup{justification = centering}
   
   {\includegraphics[width=1\columnwidth, height =22ex]{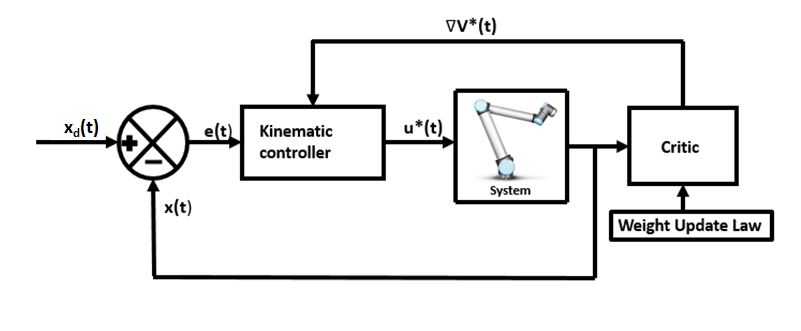}
   } 
 \vspace{-3mm}

  \caption{Block diagram of the kinematic control scheme for             \quad a robot manipulator using adaptive critic method  }

 \label{fig:Blockdiagram}
 \vspace{-3mm}
 \end{figure} In \cite{patchaikani2012single}, the kinematic control of a robot manipulator using this SNAC framework was presented. The block diagram of the kinematic control scheme for a robot manipulator using SNAC is shown in Fig. \ref{fig:Blockdiagram}. 
As these schemes used back-propagation for critic weight updates,  after repeated training, one could show the convergence to near optimal cost through extensive simulations. There exists very few works in the literature that has shown the convergence to the optimal cost in an analytical manner along with the proof of stability. In this work, we are solving the kinematic control problem of any $n$-DOF robot manipulator where the tasks of reaching a fixed target position and following a time varying task space trajectory have been solved in the framework of optimal regulation and optimal tracking respectively using SNAC. A simple and novel critic weight update rule which ensures that the closed loop  system is stable has been proposed. The optimal kinematic control policy using HJB formulation has been developed in the framework of optimal regulation and optimal tracking. Former make the robot to reach fixed target position and later make the robot to follow the task-space time varying trajectory. The analytical proof of stability and convergence has been done using Lyapunov approach. As the degrees of freedom of robotic systems are increasing, learning based strategies will play more important role as compared to model based analytics \cite{christoph2014},\cite{Ren2014}. Thus the proposed approach has significant relevance in this context. The relevant literature is further scrutinized.

The optimization based kinematic control has been accomplished using local optimization for finding an instantaneous optimal solution \cite{sqp_icra17,hollerbach1985redundancy}. It is centered on the Jacobian pseudo-inverse and null space. The redundancy is resolved by including some constraints into the direct kinematic model or by projecting a particular solution onto the Jacobian null space. The other approach is a global optimization which uses an integral type performance index along the whole trajectory \cite{kim94,hollerbach89}. The redundancy resolution is converted to an optimal control problem with the necessary conditions of optimality given by the Pontryagin's principle or by the optimal control theory. Both
of the aforementioned methods are suboptimal and require
the pseudo-inversion of Jacobian continuously over time which is computationally expensive and suffers from local instability problems \cite{maciejewski1989kinetic,o2002divergence,cocuzza2011novel}.

Existing approach towards following a time varying task space trajectory optimally is to find the feedforward term using the dynamics inversion concept and the feedback term by solving an HJB equation \cite{kiumarsi2015actor}. However, such solution is only near optimal because of the feedforward term.
Therefore by using an augmented system dynamics consisting of feedback tracking error and reference trajectory and using it to solve the HJB equation 
results in an optimal control law which is a combination of feedforward and feedback control inputs \cite{wang2018neural}.

The remainder of this paper is organised as follows. Problem formulation for optimal kinematic control is presented in Section II. In Section III and IV detailed mathematical derivation for the control scheme along with stability proof is presented. In Section V, various simulation and experimental results for kinematic motion control of a 6-DOF robotic manipulator are evaluated with comparison with the state-of-the-art kinematic control solutions. This paper is finally concluded
in Section VI.

\section{Problem Formulation}
The forward kinematics of a manipulator involves a non-linear transformation from Joint space to Cartesian space as described by:
				\begin{equation}
         	       x(t)=f(\theta(t))
                \end{equation}
where, $x(t)\in\mathbb{R}^{n\times 1}$ describes the position and orientation of end effector in workspace at time $t$, $\theta(t)\in\mathbb{R}^{m\times 1}$ each element of which describes the joint angle in the joint space at time $t$, and $f(.)$ is the non linear mapping. Because of non-linearity and redundancy of mapping, it is usually difficult to directly get $\theta(t)$ for desired $x(t)=x_{d}(t)$, where $x_{d}(t)$ is the desired end effector position. By contrast, the mapping from joint space to cartesian space at velocity level is affine mapping. Taking time derivative on both sides of (1) gives:            
             \begin{equation}
                \dot{x}(t)=J\dot{\theta}(t)
             \end{equation}
where, $J=\frac{\partial f}{\partial\theta}\in\mathbb{R}^{n\times m}$ is the Jacobian matrix of $f(.)$. 

In this paper we focus on design of angular speed of the manipulator which serves as an input to the tracking control loop for robot control. Then the robot manipulator kinematics can be rewritten by replacing $\dot{\theta}(t)$ by $u(t)$:
			\begin{equation}
				\dot{x}(t)=Ju(t)
			\end{equation}
Thus, we get the dynamics of the system as above.
In this paper we propose to solve for an optimal control input ${u}(t)$ for a robot manipulator described by (3), such that the tracking error given by,
\begin{equation}
{e}(t) = x(t) - x_d(t)
\end{equation}
for a given reference trajectory $x_d(t)$, reduces to zero with time.

\section{Optimal Regulation}

In this section, we address the optimal kinematic control of a robot manipulator, where the robot end effector has to reach a fixed target position. This is framed as an optimal regulation problem. In the context of kinematic control, optimal regulation means that the robot is made to reach a fixed target position following an optimal trajectory. This trajectory is generated by actuating the desired optimal control policy.

Let $x_{d}(t)\in\mathbb{R}^{n\times 1}$ be the  desired end effector position. 
Differentiating (4) with respect to time, error dynamics of the system is obtained as,
\begin{equation}
\begin{aligned}
\dot{e}=\dot{x}(t)
=Ju(t)
\end{aligned}
\end{equation}
\subsection{Formulation of control policy}
The infinite horizon HJB cost function for (5) is given by,
\begin{equation}
V(e(t))=\int_t^\infty{e(t)^T{Q}e(t)+u(t)^TRu(t)}dt   
\end{equation}
where, $Q\in\mathbb{R}^{n\times n}$ and $R\in\mathbb{R}^{m\times m}$ are positive definite design matrices. The control input needs to be admissible so that the cost function equation (6) is finite. The Hamiltonian for the this cost function with the admissible control input $u(t)$ is,
\begin{equation}
\begin{aligned}
  H(e(t),u(t),\nabla V(t))=&\nabla {V^*}^T\dot{e}(t)+e(t)^TQe(t)\\
  &+u(t)^TRu(t)
\end{aligned}
\end{equation}
Here, $\nabla {V^*}$ is the gradient of the cost function $V(e(t))$ with respect to $e(t)$. The optimal control input which minimizes the cost function (6) also minimizes the Hamiltonian (7). Therefore, the optimal control is found by solving the following condition $\frac{\partial H(e,u,\nabla V)}{\partial u}=0$ and obtained to be
\begin{equation}
u^*(e)=\frac{-1}{2}R^{-1}J^T\nabla {V^*}(e)
\end{equation}
Then, HJB equation may be re-written as follows,
				\begin{equation}
                0={\nabla V^*}^T\dot{e}(t)+e(t)^T{Q}e(t)+{u^*(t)}^TRu^*(t)
                \end{equation}
                with the cost function $V^*(0)=0$.
                %It is observed that $V^*$ and $\nabla V^*$ are unavailable usually. In the next section, a neural network is used to approximate the cost function.
                
\subsection{Neural Network control design}
By using the universal approximation property, $V^*(e(t))$ is constructed by a single hidden layer neural network with a non-linear activation function:
		 \begin{equation}
         V^*(e(t))=W_{c}^T\sigma_{c}(e)+\varepsilon_{c}
         \end{equation}
where, $W_{c}\in\mathbb{R}^{l}$, is the constant target NN weight vector, $\sigma_{c}\in\mathbb{R}^{l}$ is the activation function output, $l$ is the number of hidden neurons and $\varepsilon_{c} $ is the function reconstruction error. The target NN vector and reconstruction errors are assumed to be upper bounded according to $\parallel W_{c} \parallel\leq\lambda_{W_{c}}$ and $\parallel \varepsilon_{c} \parallel\leq\lambda_{\varepsilon}$. 
			\begin{equation}
        \nabla V^*=\nabla \sigma_{c}^TW_{c} + \nabla \varepsilon_{c}
        \end{equation}
Approximate optimal cost function and its gradient is given by
			\begin{align}
       \hat{V}^*=\hat{W_{c}}^T\sigma_{c}(e)\\
              \nabla\hat{V}^*=\nabla\sigma_{c}^T\hat{W_{c}}
       \end{align}
		where, $\hat{W_{c}}$ is the NN estimate of target weight vector $W_{c}$.
%         \begin{equation}
%        \nabla\hat{V}^*=\nabla\sigma_{c}^T\hat{W_{c}}
%        \end{equation}
Using (8) and (11), the optimal control law can be written as:
			\begin{equation}
      	u^*(e)=\frac{-1}{2}R^{-1}J^T(\nabla\sigma_{C}^TW{c}+\nabla \varepsilon_{c})
      		\end{equation}
Considering (8) and (13), estimated optimal control law is:
\begin{equation}\label{eq:reg_law}
\hat{u}^*(e)=\frac{-1}{2}R^{-1}J^T\nabla\sigma_{c}^T\hat{W_{c}}
\end{equation}
The modified error dynamics using (15) is,
\begin{equation}
\dot{e}=-\frac{1}{2}JR^{-1}J^T\nabla\sigma_{c}^T\hat{W_{c}}
\end{equation}

Before proposing the critic weight tuning law and stability proofs, following assumptions need to be made:
% \newline

\textit{Assumption 1:} For the system represented by its error dynamics (5), with cost function (6), let $J_{s}(e)$ be a continuously differentiable Lyapunov function candidate satisfying,
\begin{equation}
\dot{J_{s}(e)}=\nabla J_{s}(e)^T\dot{e}<0
\end{equation}
Then, there exists a positive definite matrix, $M\in\mathbb{R}^{n\times n}$ ensuring,
\begin{equation}
\nabla J_{s}(e)^T\dot{e}=-\nabla J_{s}(e)^TM\nabla J_{s}(e)
\newline\leq {-\lambda_{min}(M)}\parallel\nabla J_{s}(e) \parallel^2
\end{equation}

During the implementation, $J_{s}(e)$ can be obtained by selecting a polynomial with respect to the vector $e$, such as $J_{s}(e)=\frac{1}{2}e^Te$.
\newline
\textit{Remark 1}: Based on the result of \cite{dierks2010optimal}, the closed loop dynamics with optimal control law can be bounded by a function of the system state. In such situations, it can be asssumed that
$\parallel F(e)+G(e)(u^*) \parallel\leq\eta\parallel\nabla J_{s}(e) \parallel$ with $\eta>0$, 
$\parallel\nabla J_{s}(e)\parallel^T\parallel F(e)+G(e)(u^*)\parallel\leq\eta\parallel\nabla J_{s}(e)\parallel^2$.
Combining (17) with the fact given, $\lambda_{min}\parallel\nabla J_{s}(e) \parallel^2\leq\nabla J_{s}(e)^TM\nabla J_{s}(e)\leq\lambda_{max}(M)\parallel\nabla J_{s}(e) \parallel^2$,  it implies that \textit{assumption 1} holds.
% \newline
% \newline

Moving on, a simple and novel critic weight tuning law is proposed. 
\begin{equation}
\dot{\hat{W_{c}}}=\alpha\nabla\sigma_{c}JR^{-1}J^T\nabla J_{s}(e)
\end{equation}
where, $\alpha$ is the learning rate of critic network and $J_{s}(e)=\frac{1}{2}e^Te$.By using this weight tuning law, stability and performance is guaranteed theoretically using the Lyapunov approach.
% \newline
% \newline
\subsection{Stability Analysis}
In this section, the stability of the system for optimal regulation is investigated.
\newline
\textit{Assumption 2}: The Jacobian matrix is bounded as $\parallel J(\theta)\parallel\leq\lambda_{J}$,where $\lambda_{J}$ is a positive constant. Here, the control coefficient matrix is the Jacobian matrix. Also, $\nabla\sigma_{c},\nabla\varepsilon_{c}$ are bounded as $\parallel\nabla\sigma_{c}\parallel\leq\lambda_{\sigma} $ and $\parallel\nabla\varepsilon_{c}\parallel\leq\lambda_{\nabla\varepsilon}$ where $\lambda_{\sigma}$ and $\lambda_{\nabla\varepsilon}$ are positive constants.
\newline
% Now, let the stability proof is presented using the Lyapunov approach,
% \newline
The Lyapunov candidate function is selected as, \begin{equation}
L(t)=\frac{1}{4\alpha}\widetilde{W_{c}}^T\widetilde{W_{c}}+J_{s}(e)
\end{equation}
 \begin{equation}
\dot{L}(t)=\frac{1}{2\alpha}\widetilde{W_{c}}^T\dot{\widetilde{W_{c}}}+\nabla J_{s}^T\dot{e}
\end{equation} where, $\widetilde{W}=W_{c}-\hat{W_{c}}$. Using (19),
\begin{equation}
\dot{L}(t)=\frac{-1}{2}\widetilde{W_{c}}^T\nabla\sigma JR^{-1}J^T\nabla J_{s}+\nabla J_{s}^T(\dot{e})
\end{equation}
The system error dynamics of the system for the optimal control law is $\dot{e}^*=J{u}^*$. Using the control laws (14) and (15),
% \begin{equation}
\begin{align}
\dot{e}^*-\dot{e}=& J(u^*-\hat{u}^*)
				= -\frac{1}{2}JR^{-1}J^T(\nabla\sigma_{c}^T\widetilde{W_{c}}+\nabla\varepsilon_{c})
\end{align}
% \end{equation}
\vspace{-3mm}

Using (23) in (22) and on simplification,
% \begin{equation}
\begin{align}
\dot{L}(t)=\nabla J_{s}^T\dot{e}^*+\frac{1}{2}\nabla J_{s}^TJR^{-1}J^T\nabla\varepsilon_{c}
\end{align}
% \end{equation}
Applying \textit{Assumption 1} and \textit{Assumption 2} here
\begin{align}
\dot{L}\leq {-}\lambda_{min}(M)\parallel\nabla J_{s}\parallel^2+\frac{1}{2}\parallel\nabla J_{s}\parallel\lambda_{J}^2\parallel R^{-1}\parallel\lambda_{\nabla\varepsilon}
\end{align}
From (25), the following inequality may be derived:\vspace{-1mm}
\begin{align}
\parallel\nabla J_{s}\parallel\geq \frac{1}{2\lambda_{min}(M)}\lambda_{J}^2\parallel R^{-1}\parallel\lambda_{\nabla\varepsilon}
\end{align}
With the above condition to be true, $\dot{L}\leq 0$ and the system is stable in the sense of Lyapunov implying that $\widetilde{W_c}$ and $e$ are both bounded. It may be expressed as :
% \newline
$\parallel\widetilde{W_c}\parallel\leq\lambda_{\widetilde{W_c}}$.

It may be also observed that the estimated optimal controller (15) converges to a neighborhood of the optimal feedback controller (14) with a finite bound $\lambda_{u}$ as,
\begin{equation}
\parallel u^*(e)-\hat{u}^*(e)\parallel\leq\frac{1}{2}\parallel R^{-1}\parallel\lambda_{J}(\lambda_{\sigma}\lambda_{\widetilde{W_{c}}}+\lambda_{\nabla\varepsilon})\overset{\Delta}{=}\lambda_{u}
\end{equation}
Hence, it may be concluded that the instantaneous cost function $V(e,u,t)=e(t)^TQe(t)+u(t)^TRu(t)$ is also bounded.

Taking time derivative of (24),
\begin{equation}
\begin{aligned}
\ddot{L}=\dot{e}^T\dot{e}^*+e^T\ddot{e}^*+\frac{1}{2}\dot{e}^TJR^{-1}J^T\nabla\varepsilon_{c}+e^T\dot{J}R^{-1}J^T\nabla \varepsilon_{c}+\\
\frac{1}{2}e^TJR^{-1}J^T\nabla^2\varepsilon_{c}\dot{e}
\end{aligned}
\end{equation}
\textit{Remark 2:}
\newline
All the terms in $\ddot{L}$ can be shown to be bounded.

It may be observed that $\ddot{L}$ is bounded and Barbalat's Lemma \cite{vamvoudakis2016control} can be invoked to conclude the asymtotic stability of the system and convergence of the parameter estimation error and the weight estimation errors towards zero. In other words, it ensures that $\widetilde{W_{c}}\rightarrow 0$.%\vspace{-2.5mm}
\section{Optimal Tracking Control}
In this section, the kinematic control of a robot manipulator following a time varying task space trajectory is solved in the framework of optimal tracking.

\subsection{Formation of Augmented System}

Let the time varying reference trajectory, denoted by $x_{d}(t)\in\mathbb{R}^{n}$ be possessing the dynamics
\begin{align}
                \dot{x_{d}}(t)=\varphi(x_{d}(t))
             \end{align}
             with the initial condition $x_{d}(0)={x_{d}}_{0}$, where, $\varphi(x_{d}(t))$ is a Lipschitz continuous function satisfying $\varphi(0)=0$. Let the trajectory tracking error be $e(t)=x(t)-x_{d}(t)$ with the initial condition $e(0)=e_{0}=x_{0}-{x_{d}}_{0}$. Considering (3) and (29), the tracking error dynamics is:
			%\begin{equation}
			%	\dot{z}(t)=f(z(t)+r(t)-\varphi((t))) + g(z(t)+r(t))u(t)
			%\end{equation}
            \begin{equation}
				\dot{e}(t)=-\varphi(x_{d}(t)) + Ju(t)
			\end{equation}
            
Next, an augmented system is constructed as in \cite{wang2018neural},in the form $\xi(t)=[e(t)^T,x_{d}(t)^T]^T\in\mathbb{R}^{2n}$ with initial condition $\xi(0)=\xi_{0}=[e_{0}^T,{x_{d}}_{0}^T]^T$. The augmented dynamics based on (29) and (30) can be formulated as:

\vspace{-3mm}

			\begin{equation}
				\dot{\xi}(t)=F(\xi(t)) + G(\xi(t))u(t)
			\end{equation}\vspace{-3mm}

where, F(.) and G(.) are the new system matrices.

\subsection{Formulation of Control Policy}

The infinite horizon HJB cost function for the system in (31) is given by,
\begin{equation}
 V(\xi(t))=\int_t^\infty{\xi(t)^T\bar{Q}\xi(t)+u(t)^TRu(t)}.dt                \end{equation}
where, $\bar{Q}=diag\{Q,0_{n\times n}\}$. $Q\in\mathbb{R}^{n\times n}$ and  $R\in\mathbb{R}^{m \times m}$ are positive definite matrices. The control input must be admissible so that cost function given by (32) is finite.

Applying the same methods of solving as in Section III-A, the control policy is obtained as,
\begin{equation}
u^*(\xi)=\frac{-1}{2}R^{-1}G(\xi)^T\nabla V^*(\xi)
\end{equation}

\subsection{Neural Network control design}
Here, $V^*(\xi(t))$ is constructed by a single hidden layer neural network with a non-linear activation function:
		 \begin{equation}
         V^*(\xi(t))=W_{c}^T\phi_{c}(\xi)+\varepsilon_{c}
         \end{equation}
where, $W_{c}\in\mathbb{R}^{p}$, is the ideal weight, $\parallel W_{c} \parallel\leq\lambda_{W_{c}}, \phi_{c}\in\mathbb{R}^{p}$ is the activation function output, $p$ is the number of hidden neurons and $\varepsilon_{c}$ is the function approximation error. %Taking the gradient of the cost function,

 Using the same approach as in Section III-B, optimal control law and approximate control law may be obtained as,
			\begin{equation}
      	u^*(\xi)=\frac{-1}{2}R^{-1}G(\xi)^T(\nabla\phi_{C}^TW{c}+\nabla \varepsilon_{c})
      		\end{equation}
           
\begin{equation}\label{eq:tracking_law}
\hat{u}^*(\xi)=\frac{-1}{2}R^{-1}G(\xi)^T\nabla\phi_{c}^T\hat{W_{c}}
\end{equation}
Applying (\ref{eq:tracking_law}) into the augmented system dynamics (31), $\dot{\xi}(t)=F(\xi) + G(\xi)\hat{u}^*(t)$ can be formulated as:
\begin{equation}
\dot{\xi}=F(\xi)-\frac{1}{2}G(\xi)R^{-1}G(\xi)^T\nabla\phi_{c}^T\hat{W_{c}}
\end{equation}

The weight update law proposed for the critic of optimal tracking control is,

\begin{equation}
\dot{\hat{W_{c}}}=\alpha\nabla\phi_{c}G(\xi)R^{-1}G(\xi)^T\nabla J_{s}(\xi)
\end{equation}
where, $\alpha$ is the learning rate of critic network and $J_{s}(\xi)=\frac{1}{2}\xi^T\xi$.

\textit{Remark 3:} \textit{Assumption 1} and \textit{Remark 1 } holds here. The augmented system states shall replace the system error states as follows ; For the augmented system (31), with cost function (32), let $J_{s}(\xi)$ be a continuously differentiable Lyapunov function candidate satisfying,
\begin{equation}
\dot{J}_{s}(\xi)=\nabla J_{s}(\xi)^T\dot{\xi}<0
\end{equation}
Then,there exists a positive definite matrix, $M\in\mathbb{R}^{2n\times 2n}$ensuring,
\begin{equation}
\nabla J_{s}(\xi)^T\dot{\xi}=-\nabla J_{s}(\xi)^TM\nabla J_{s}(\xi)
\newline\leq {-\lambda_{min}(M)}\parallel\nabla J_{s}(\xi) \parallel^2
\end{equation}
During the implementation,$J_{s}(\xi)=\frac{1}{2}\xi^T\xi$.

\subsection{Stability Analysis}

In this section, the stability of the system is investigated.

\textit{Remark 4}: \textit{Assumption 2} holds here. However, the control coefficient matrix is modified. The assumption $\parallel G(\xi)\parallel\leq\lambda_{g}$, where $\lambda_{g}$ is a positive constant shall be used.
\newline
Lyapunov function, \begin{equation}
L(t)=\frac{1}{4\alpha}\widetilde{W_{c}}^T\widetilde{W_{c}}+J_{s}(\xi)
\end{equation}
 \begin{equation}
\dot{L}(t)=\frac{1}{2\alpha}\widetilde{W_{c}}^T\dot{\widetilde{W_{c}}}+\nabla J_{s}^T\dot{\xi}
\end{equation}
Using the similar approach as in Section III-C,

\begin{equation}
\begin{aligned}
\dot{L}(t)=\nabla J_{s}^T\dot{\xi}^*+\frac{1}{2}\nabla J_{s}^TG(\xi)R^{-1}G(\xi)^T\widetilde{W_{c}}
\end{aligned}
\end{equation}
Applying the \textit{Remark 3} and \textit{Remark 4} here
\vspace{-2mm}

\begin{equation}
\dot{L}\leq {-}\lambda_{min}(M)\parallel\nabla J_{s}\parallel^2+\frac{1}{2}\parallel\nabla J_{s}\parallel\lambda_{g}^2\parallel R^{-1}\parallel\lambda_{\nabla\varepsilon}
\end{equation}
From this, we observe that if the  condition below holds, then, $\dot{L}(t)\leq 0$ is negative semidefinite implying that the system is stable in the sense of Lyapunov. 
\begin{equation}
\parallel\nabla J_{s}\parallel\geq \frac{1}{2\lambda_{min}(M)}\lambda_{g}^2\parallel R^{-1}\parallel\lambda_{\nabla\varepsilon}
\end{equation}
\vspace{-2mm}

Hence, $\widetilde{W_c}$ and $\xi$ are both bounded. It may be expressed as :
% \newline
$\parallel\widetilde{W_c}\parallel\leq\lambda_{\widetilde{W_c}}$

Also,
\begin{equation}
\parallel u^*(\xi)-\hat{u}^*(\xi)\parallel\leq\frac{1}{2}\parallel R^{-1}\parallel\lambda_{g}(\lambda_{\sigma}\lambda_{\widetilde{W_{c}}}+\lambda_{\nabla\varepsilon_{c}})\overset{\Delta}{=}\lambda_{u}
\end{equation}
It may be noted that  estimated controller input converges to the neighbourhood of the optimal control input as a finite bound $\lambda_{u}$ exists just as in the case of regulation.
It is also observed that the cost function $V(\xi,u,t)=\xi(t)^T\bar{Q}\xi(t)+u(t)^Tu(t)$ is also bounded.

\begin{figure}[!h]
 \centering
 \captionsetup{justification = centering}
%    \subfloat[]
   {\includegraphics[width=0.8\columnwidth, height =45ex]{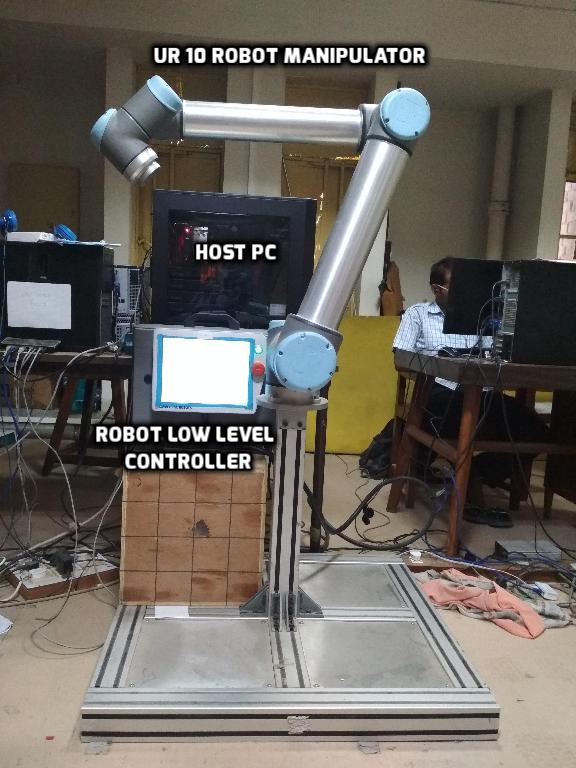}
   }
  \caption{Hardware Setup }
 \label{fig:setup}
 \end{figure}
Using the same approach as in Section III-C, it may be shown that $\ddot{L}$ is bounded. Recalling Barbalat's Lemma \cite{selmic1999backlash},
it may be concluded that the system is stable and that the parameter estimation error and weight estimation error converge to zero. %Note that the convergence of the weight estimation error ensures $\widetilde{W}\rightarrow 0$.
\vspace{-2mm}
\begin{figure*}[!ht]
 \centering
 \captionsetup{justification = centering}
   \subfloat[]
   {\includegraphics[width=0.475\columnwidth, height =18ex]{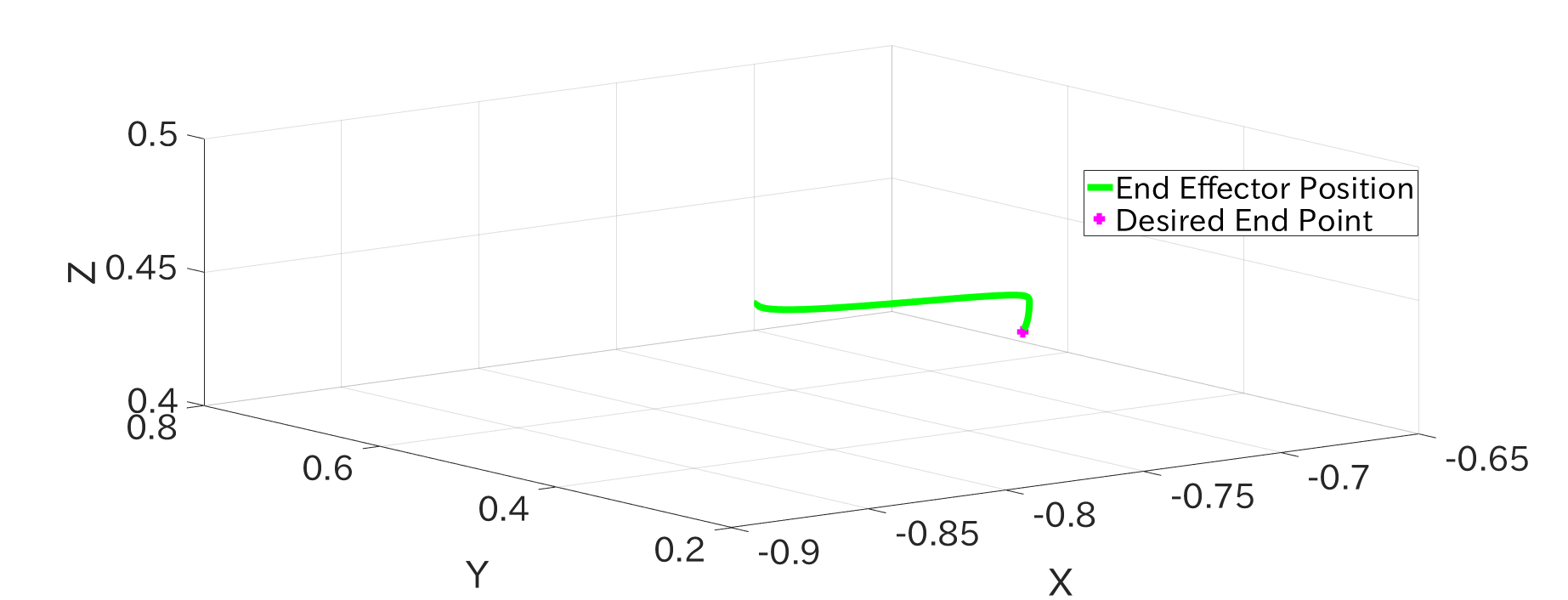}
   }
   \hfill
   \subfloat[]
   {
    \includegraphics[width=0.475\columnwidth, height =18ex]{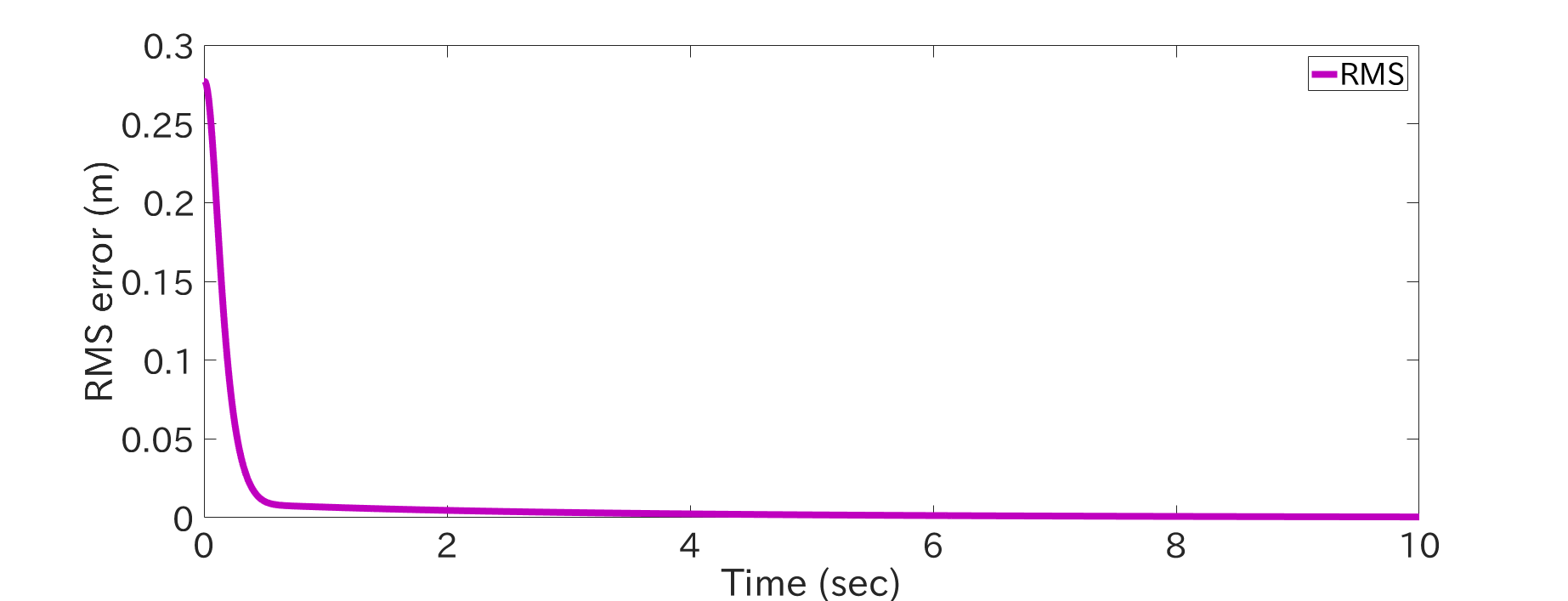}
    }  
    \hfill
    \subfloat[]
   {\includegraphics[width=0.475\columnwidth, height =18ex]{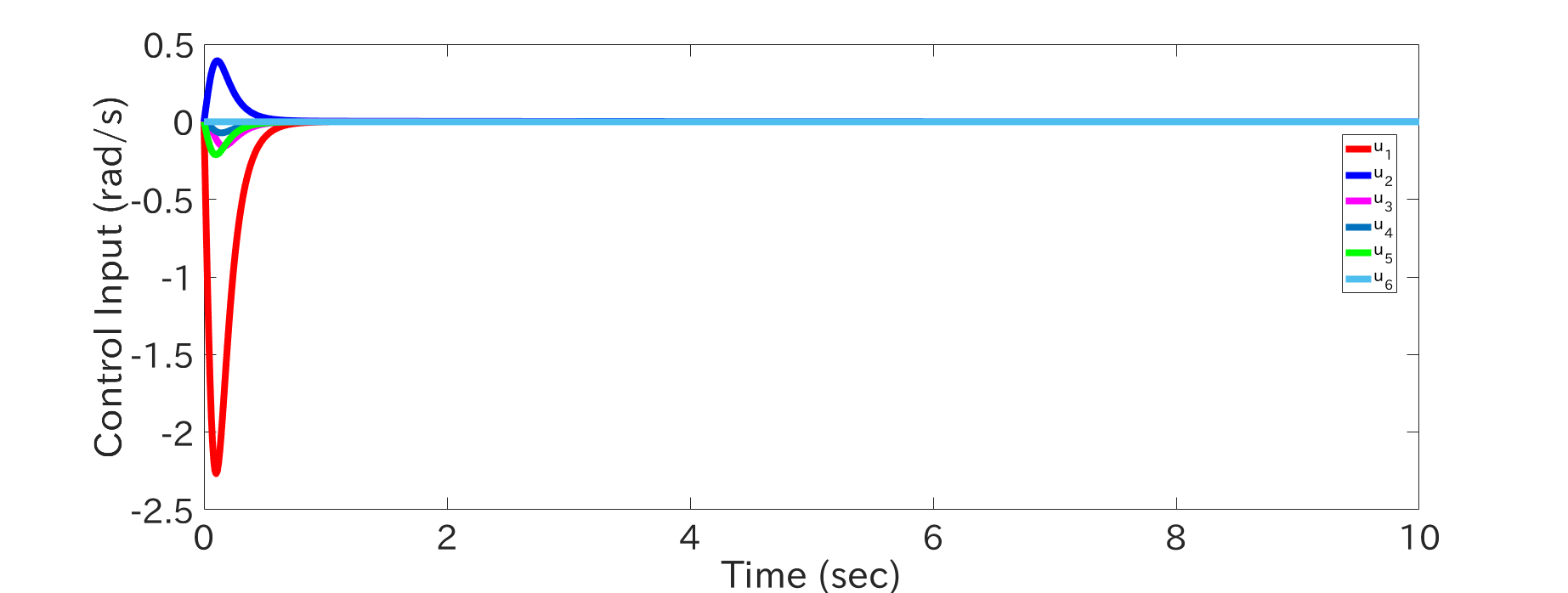}
   }
   \hfill
   \subfloat[]
   {
    \includegraphics[width=0.475\columnwidth, height =18ex]{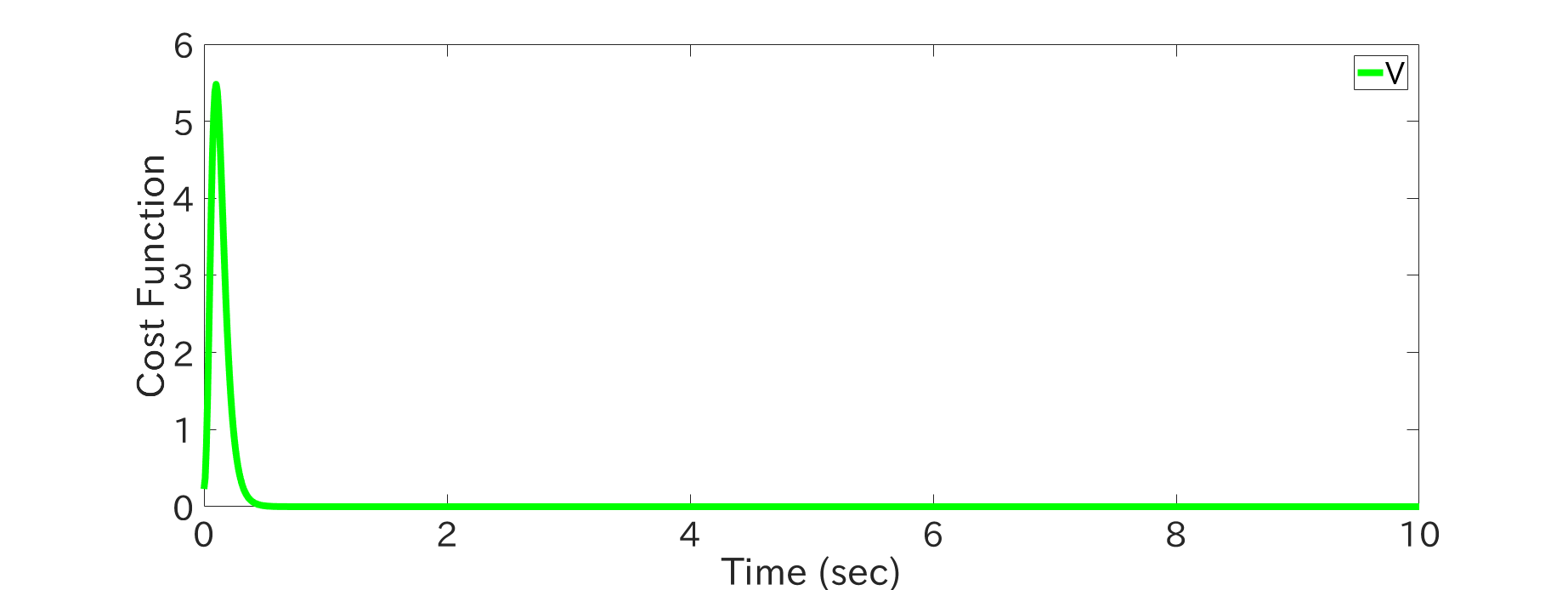}
    } \\
   \subfloat[]
   {\includegraphics[width=0.475\columnwidth, height =18ex]{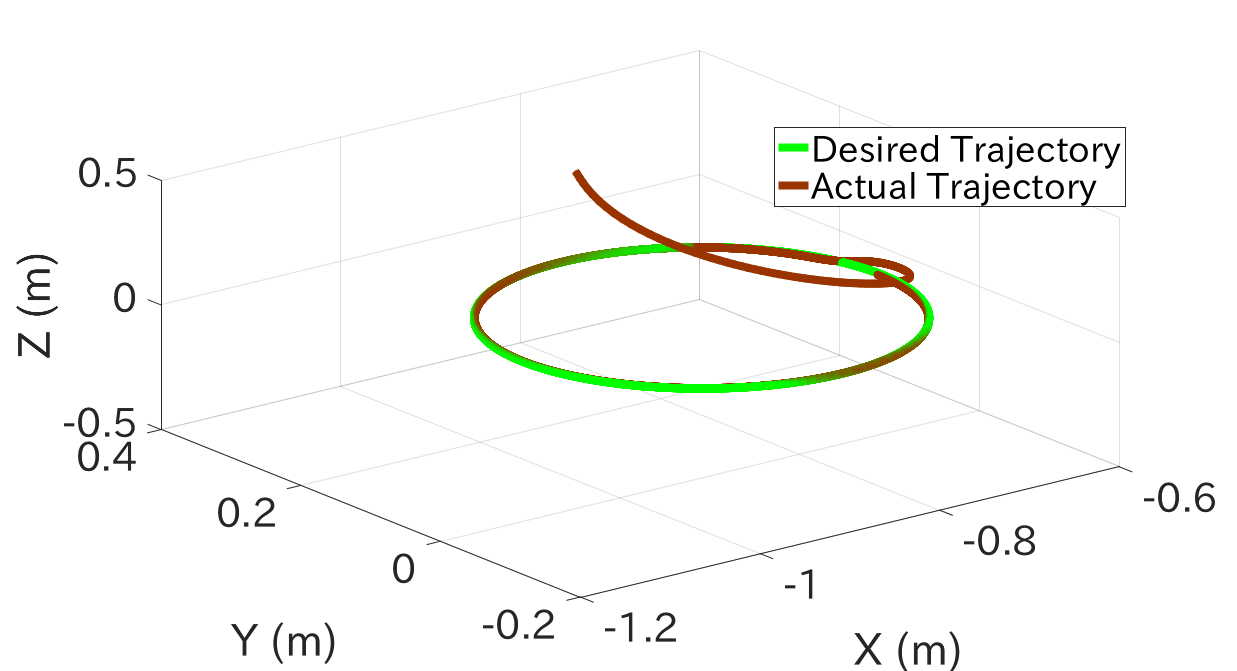}
   }
   \hfill
   \subfloat[]
   {
    \includegraphics[width=0.475\columnwidth, height =18ex]{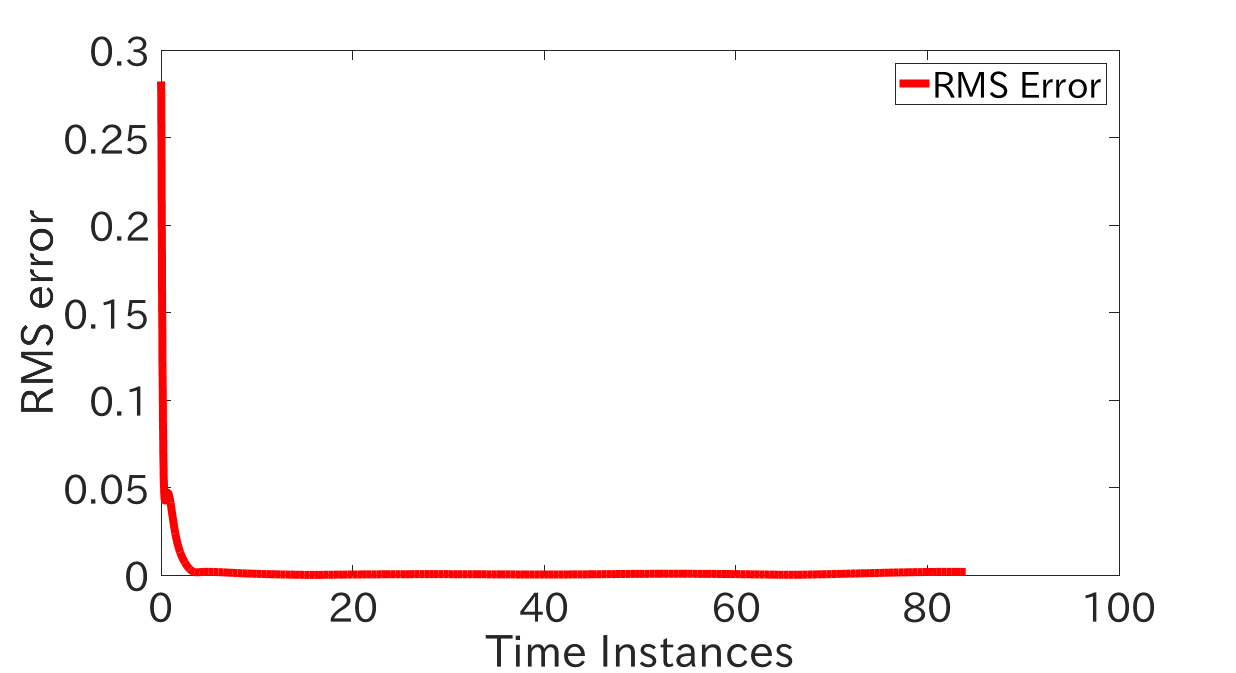}
    }  
    \hfill
    \subfloat[]
   {\includegraphics[width=0.475\columnwidth, height =18ex]{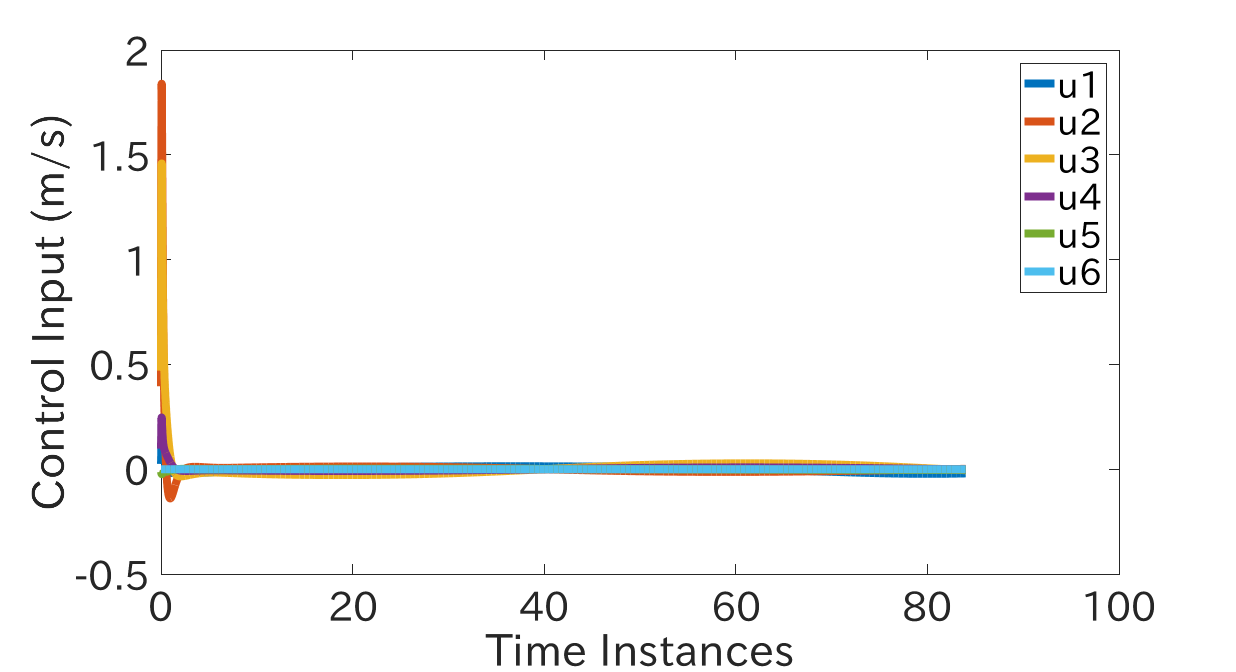}
   }
   \hfill
   \subfloat[]
   {
    \includegraphics[width=0.475\columnwidth, height =18ex]{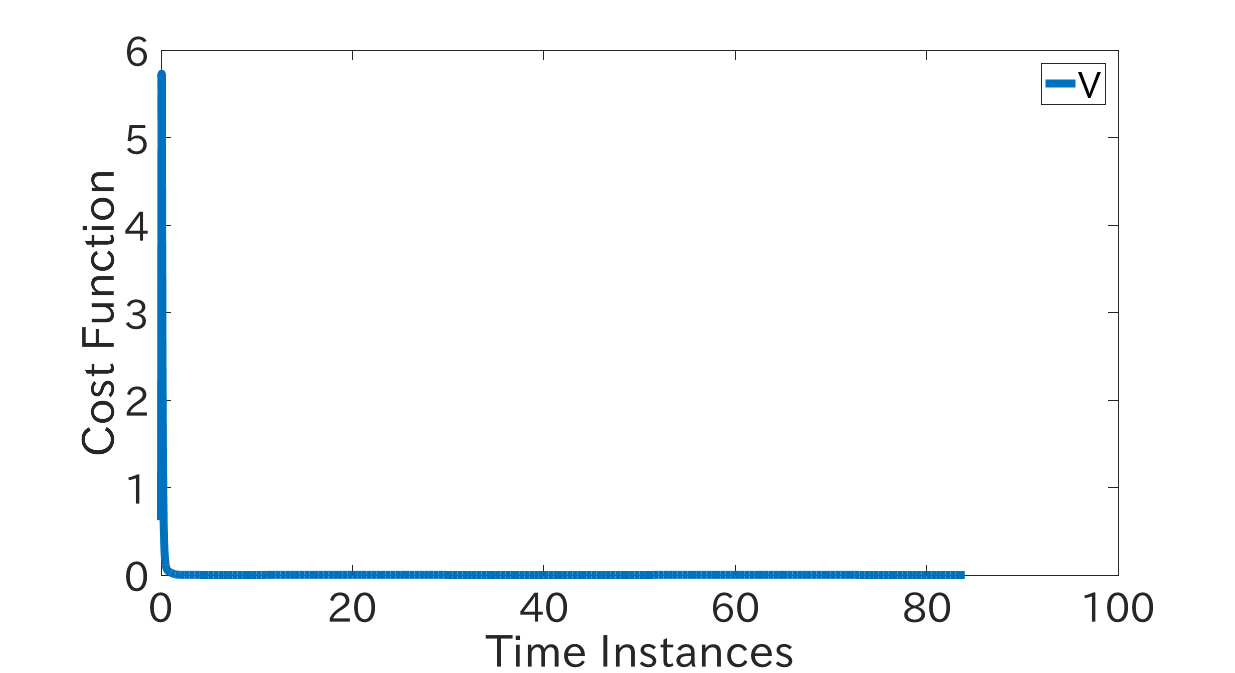}
    }

  \caption{Simulation results for Optimal regulation of a fixed target (a-d) and optimal tracking for a time varying elliptical trajectory (e-h). }
 \label{fig:sim}
 \end{figure*}
\vspace{-2mm}

 \begin{figure*}[!ht]
 \centering
 \captionsetup{justification = centering}
   \subfloat[]
   {\includegraphics[width=0.475\columnwidth, height =18ex]{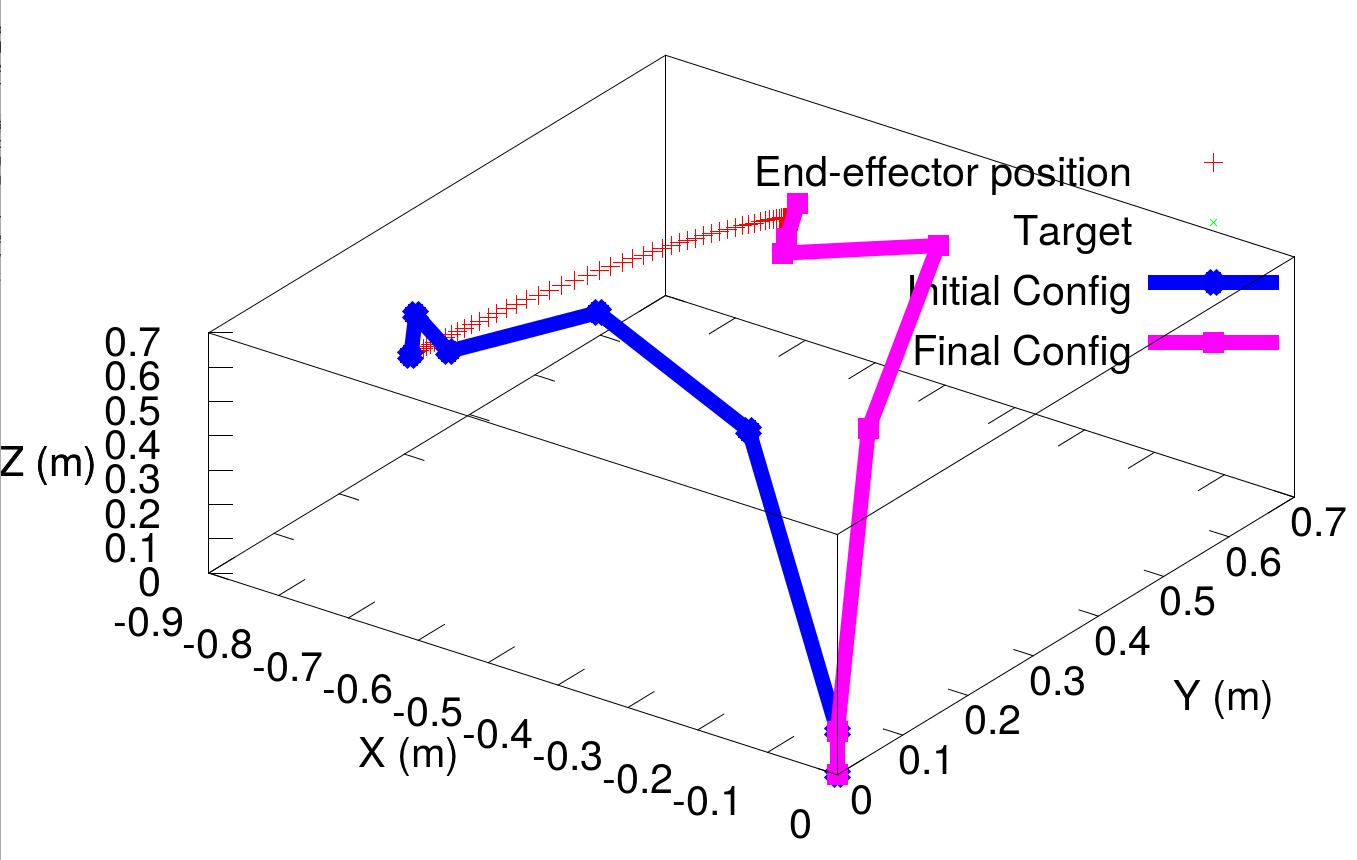}
   }
   \hfill
   \subfloat[]
   {
    \includegraphics[width=0.475\columnwidth, height =18ex]{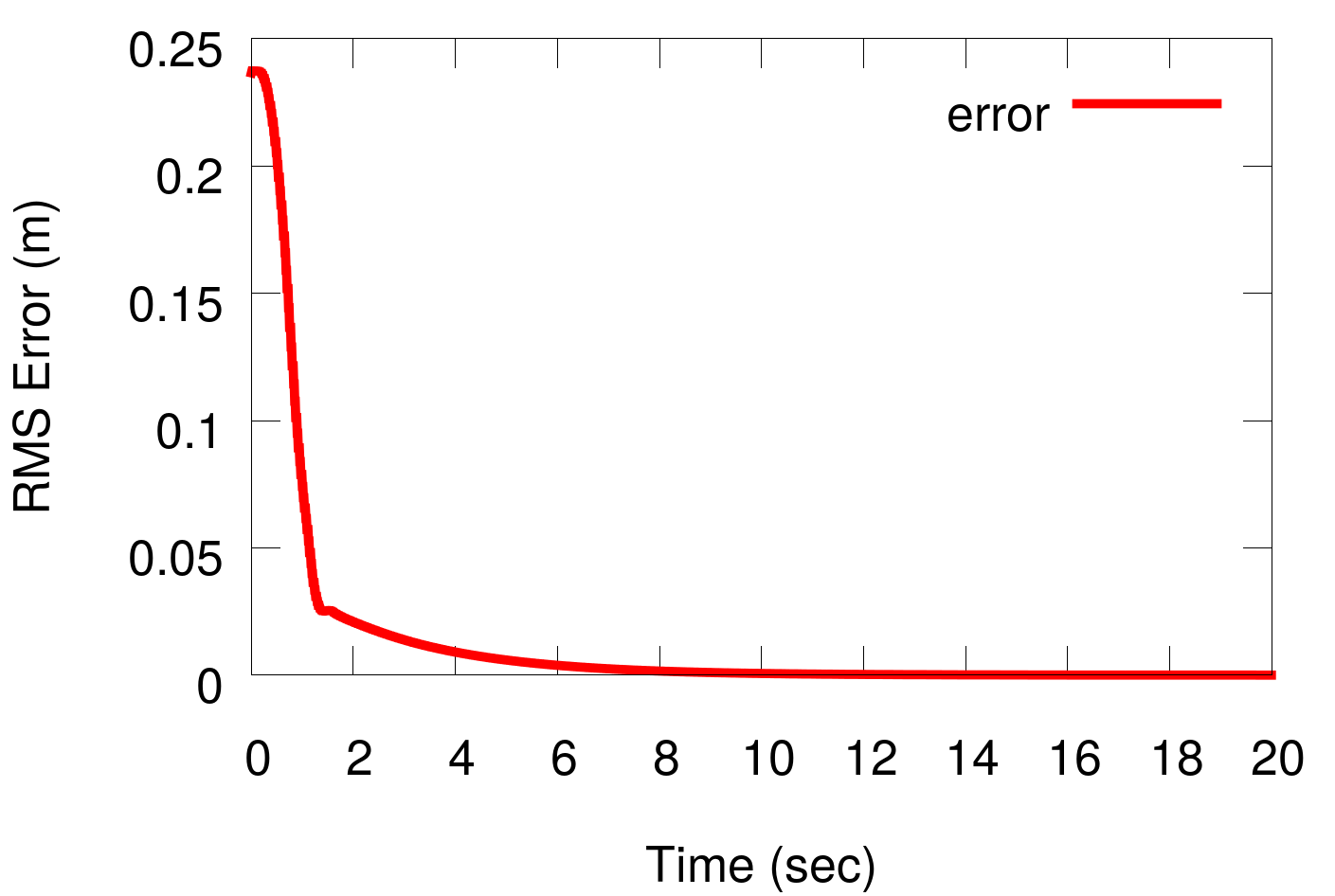}
    }  
    \hfill
    \subfloat[]
   {\includegraphics[width=0.475\columnwidth, height =18ex]{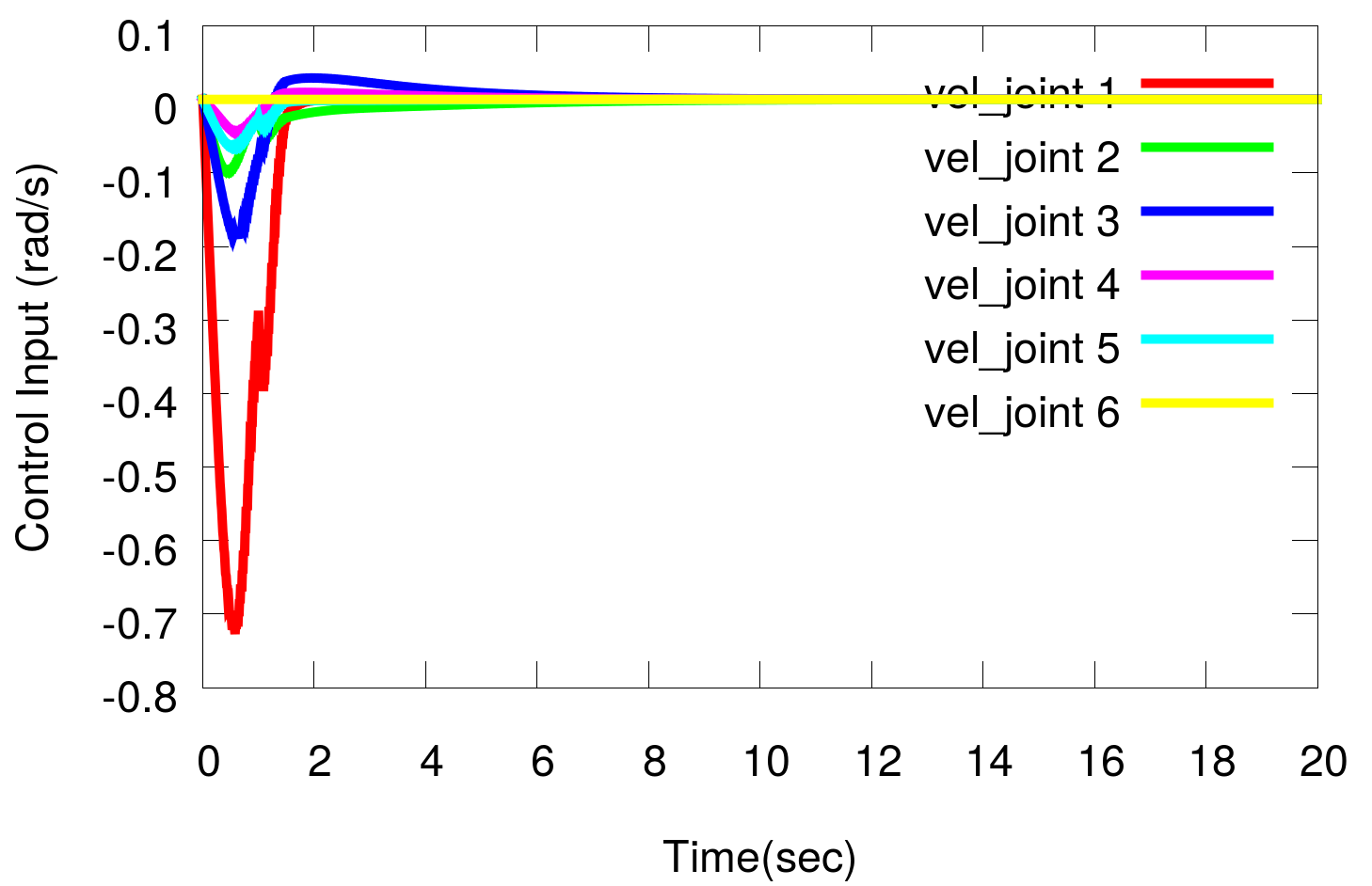}
   }
   \hfill
   \subfloat[]
   {
    \includegraphics[width=0.475\columnwidth, height =18ex]{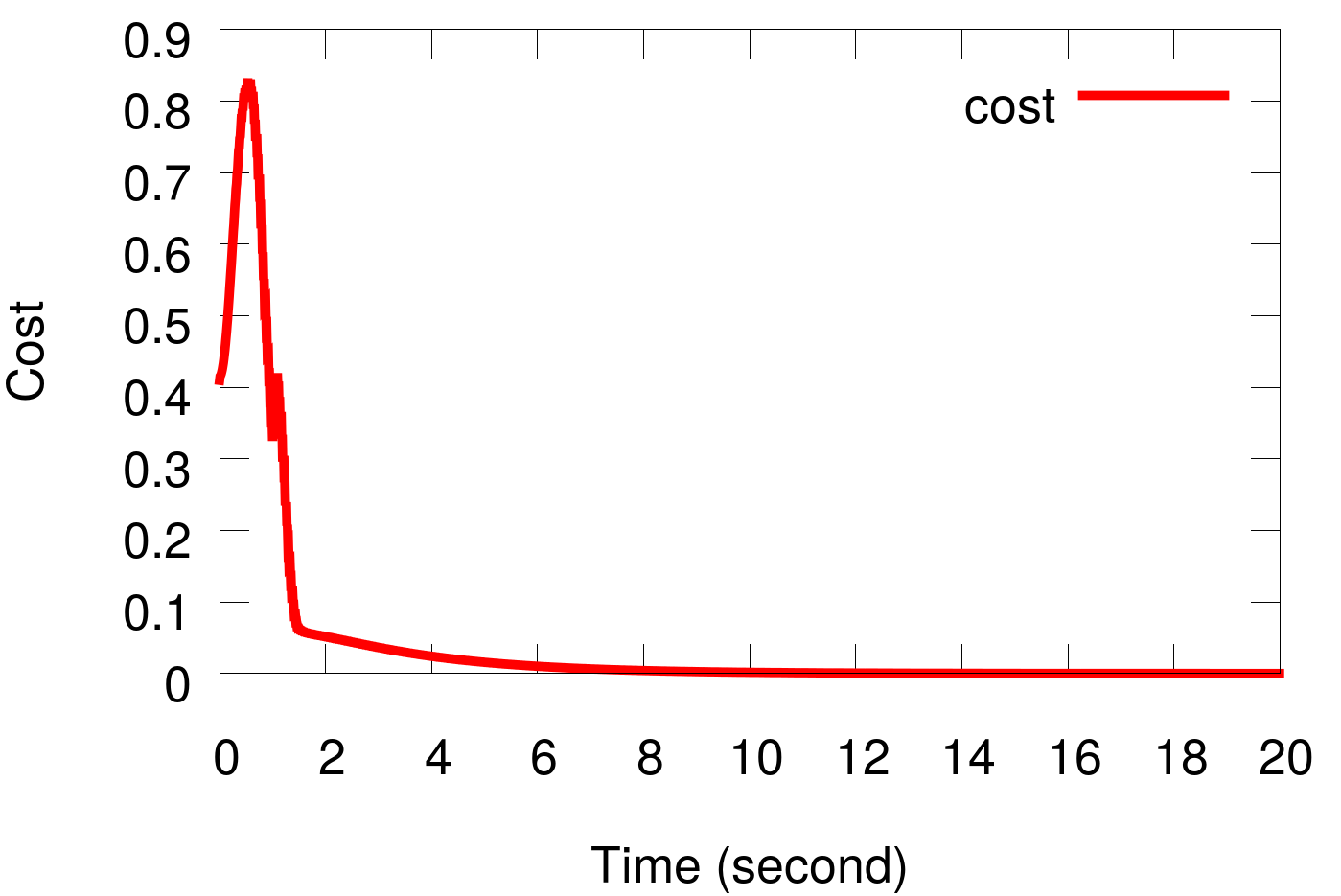}
    }     \\\vspace{-4mm}
    \subfloat[]
   {\includegraphics[width=0.475\columnwidth, height =18ex]{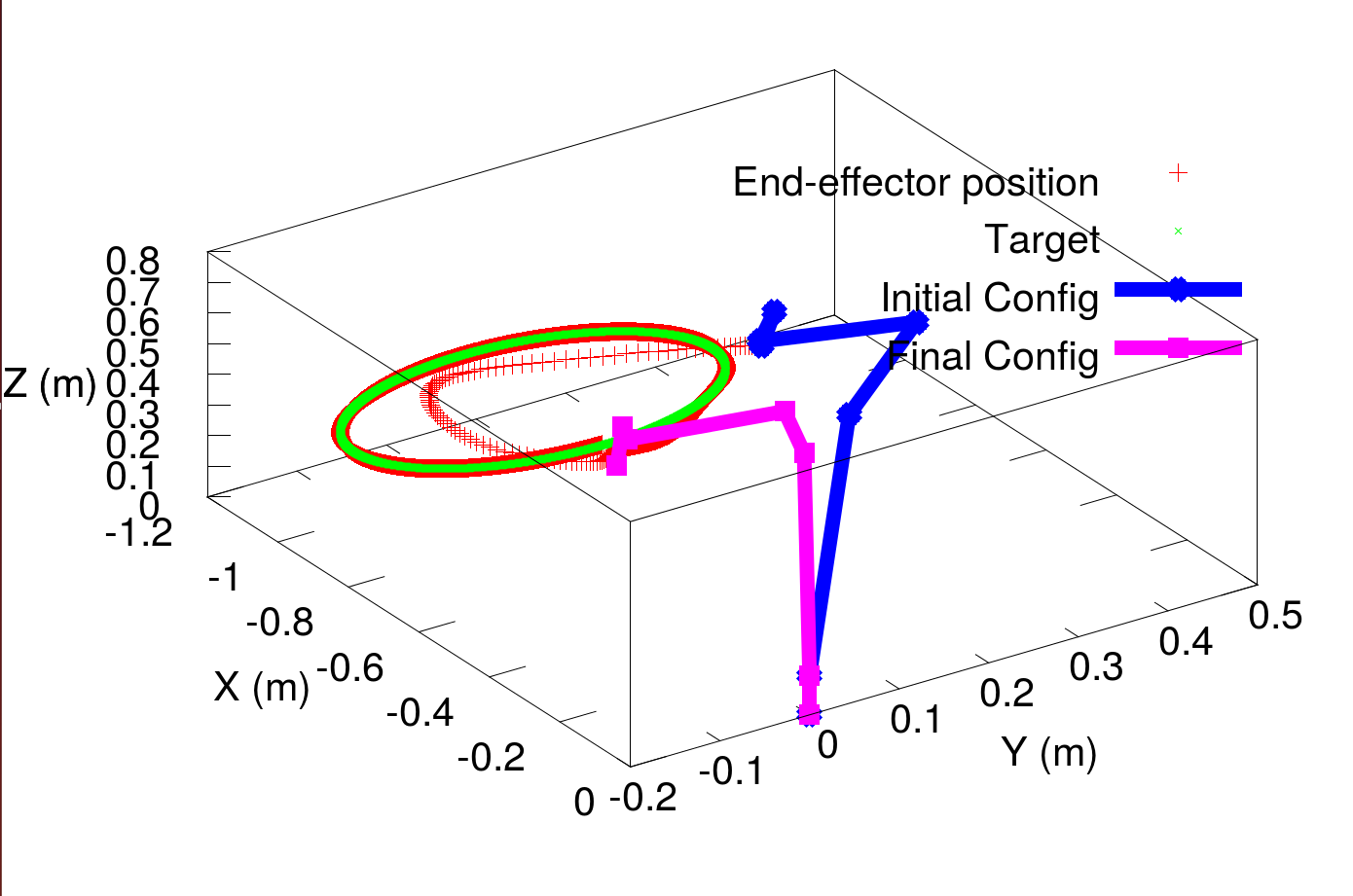}
   }
   \hfill
   \subfloat[]
   {
    \includegraphics[width=0.475\columnwidth, height =18ex]{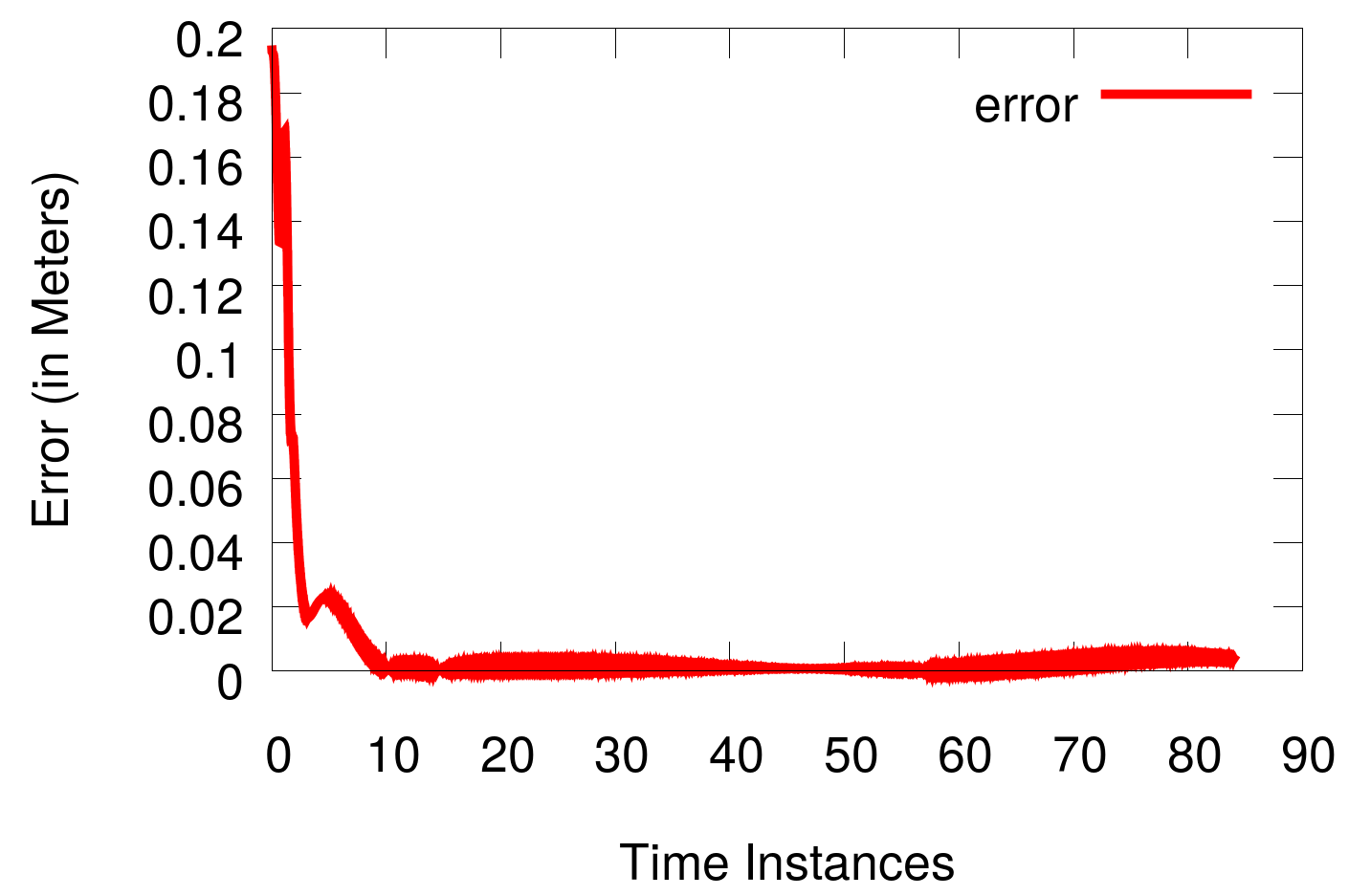}
    }  
    \hfill
    \subfloat[]
   {\includegraphics[width=0.475\columnwidth, height =18ex]{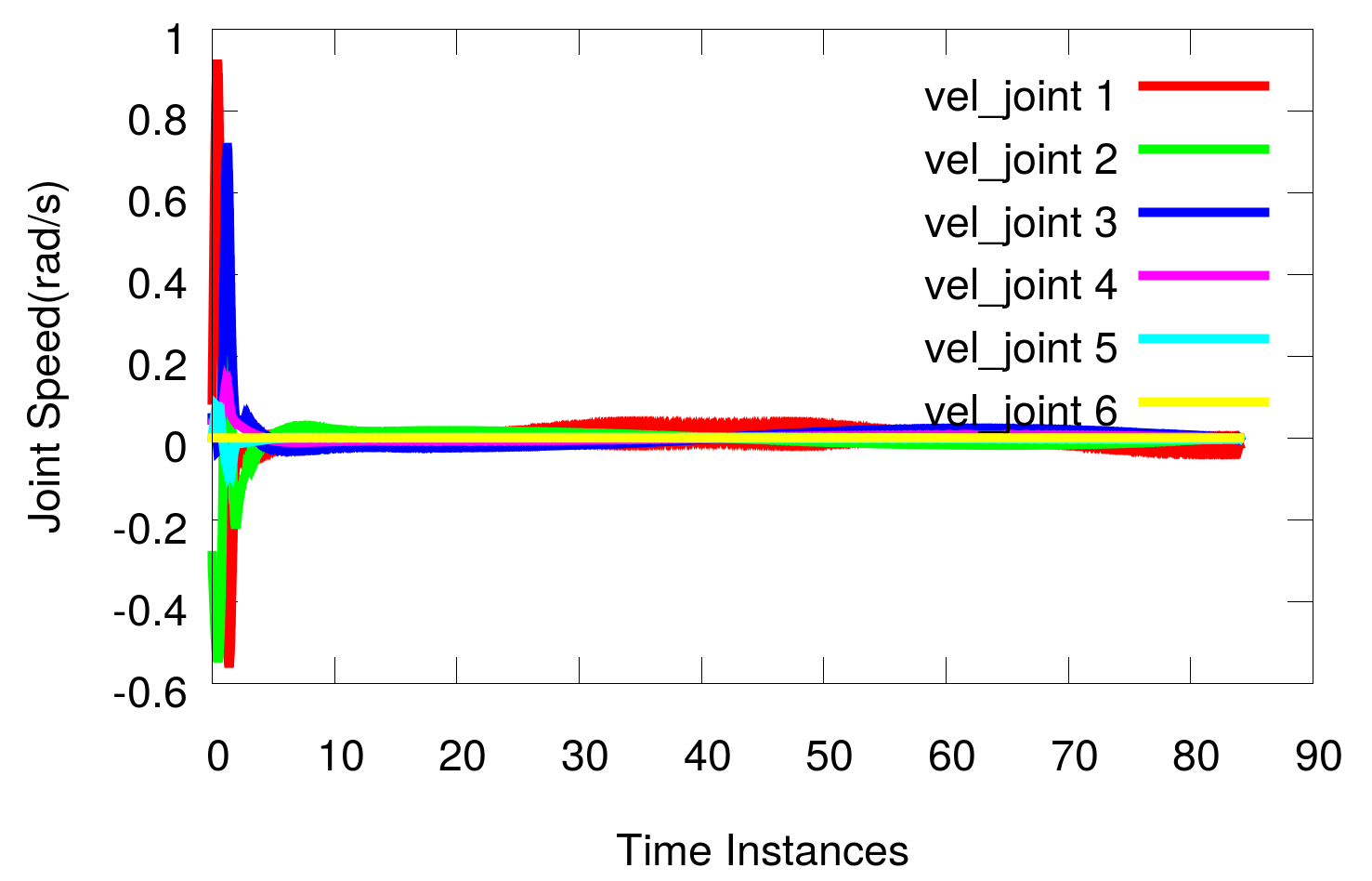}
   }
   \hfill
   \subfloat[]
   {
    \includegraphics[width=0.475\columnwidth, height =18ex]{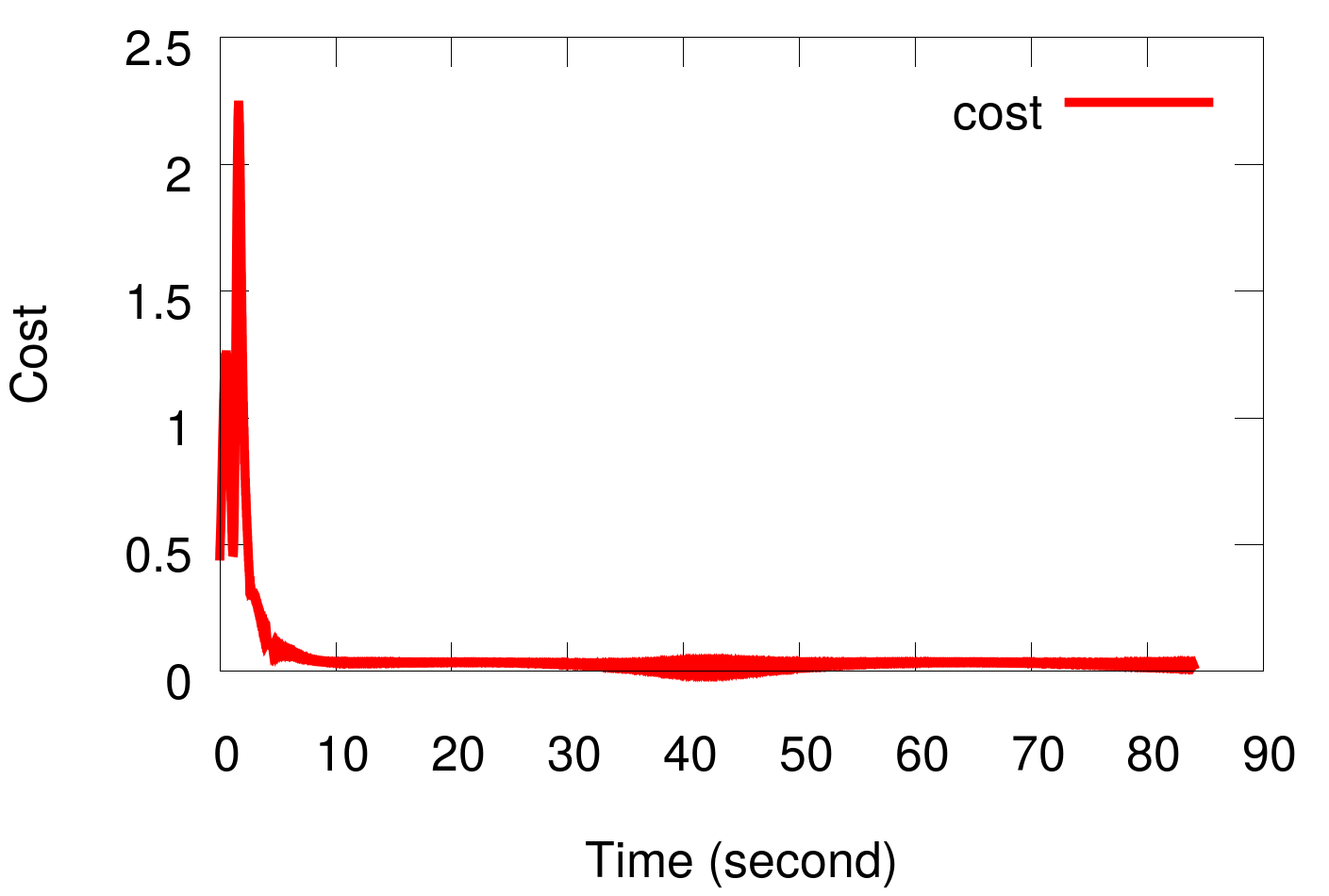}
    } 
    
  \caption{Experimental results for Optimal regulation of a fixed target (a-d) and optimal tracking for a time varying elliptical trajectory (e-h).  }
 \label{fig:exp}
 \vspace{-3mm}
 \end{figure*}

\section{Results and Discussion}

In this section, we consider the numerical simulations followed by real time experimental validations on a real 6 DOF UR 10 robot manipulator to demonstrate the effectiveness of the proposed kinematic control.
\subsection{Experimental Setup}\vspace{-2mm}
Our experimental setup shown in Fig. \ref{fig:setup} consists of a UR10 robot manipulator with its controller box/internal computer and a host PC/external computer. The UR10 robot manipulator is a 6 DOF robot arm designed to safely work alongside and in collaboration with a human. This arm can follow position commands like a traditional industrial robot, as well as take velocity commands to apply a given velocity in/around a specified axis. 
% In addition to standard programming, the robots have a freedrive mode for tactile programming. 
The low level robot controller is a program running on UR10's internal computer broadcasting robot arm data , receiving and interpreting the commands and controlling the arm accordingly. There are several options for communicating with the robot low level controller to control the robot including the teach pendent or opening a TCP socket (C++/Python) on a host computer. We used open source C++ based UrDriver wrapper class integrated with ROS on a host PC to implement our proposed velocity based kinematic control scheme. The host PC streams joint velocity commands via URScript to the robot real time interface over Ethernet at $125 Hz$. The driver was configured with necessary parameters like IP address of the robot at startup using ROS parameter server. \vspace{-2mm}
\subsection{Reaching a Fixed Position}
\vspace{-1.5mm}
Simulations are first performed to verify the kinematic control of the robot manipulator to reach a fixed position using the control law expressed in Equation (\ref{eq:reg_law}). A typical simulation run generated with a random seed pose $(\theta_{0}=[-0.51 -1.04 1.48 -1.82 -1.45 -1.62]rad)$ to a fixed target pose $[-0.658, 0.626, 0.407]m$ in cartesian space is shown in Figure \ref{fig:sim}(a).The weights were initialized randomly and $\alpha=\alpha_{initial}(tanh(n-k))+\alpha_{final}$, where, n=50, k=time instance, $\alpha_{initial}=100$ and $\alpha_{final}=150$.%The cost function is $ V(e(t))=\int_t^\infty{e(t)^TQe(t)+u(t)^TRu(t)}dt   $, where $Q=I_{n\times n}$ and $R=I_{m\times m}$.
 The predicted cost function was: 
$w_{1}e_{1}^2+w_{2}e_{1}e_{2}+w_{3}e_{1}e_{3}+w_{4}e_{2}^2+w_{5}e_{2}e_{3}+w_{6}e_{3}^2$

\subsection{Following a Time Varying Reference Trajectory}
In this section, the kinematic control of the robot manipulator to follow a time varying reference trajectory is validated. Simulations are first performed using the control law derived in Equation (\ref{eq:tracking_law}). A typical simulation generated with a random seed pose $(\theta_{0}=[-0.51 -1.04 1.48 -1.82 -1.45 -1.62]rad)$ and following a time varying reference trajectory moving at an angular speed of $0.075 \quad rad/s$ along a circle centered at $[-0.7,0,0.5]m$ with a radius of $0.2 m$ is shown in Figure \ref{fig:sim}(e). Weights %of the neural network  $w \in \mathbb{R}^{21\times 1}$%
were initialized randomly and $\alpha=\alpha_{initial}(tanh(n-k))+\alpha_{final}$, where, n=10, k=time intance, $\alpha_{initial}=20$ and $\alpha_{final}=70$. %The cost function is $V(\xi(t))=\int_t^\infty{\xi(t)^T\bar{Q}\xi(t)+u(t)^TRu(t)}dt   $, where $\bar{Q}=diag(I_{n\times n,}0_{n\times n})$ and $R=I_{m\times m}$.  \newline
 In this work, we take the predicted cost to be: 

$w_{1}\xi_{1}^2+w_{2}\xi_{1}\xi_{2}+w_{3}\xi_{1}\xi_{3}+w_{4}\xi_{1}\xi_{4}+w_{5}\xi_{1}\xi_{5}+w_{6}\xi_{1}\xi_{6}+w_{7}\xi_{2}^2+w_{8}\xi_{2}\xi_{3}+w_{9}\xi_{2}\xi_{4}+w_{10}\xi_{2}\xi_{5}+w_{11}\xi_{2}\xi_{6}+w_{12}\xi_{3}^2+w_{13}\xi_{3}\xi_{4}+w_{14}\xi_{3}\xi_{5}+w_{15}\xi_{3}\xi_{6}+w_{16}\xi_{4}^2+w_{17}\xi_{4}\xi_{5}+w_{18}\xi_{4}\xi_{6}+w_{19}\xi_{5}^2+w_{20}\xi_{5}\xi_{6}+w_{21}\xi_{6}^2.$

\subsection{Observations}
After a short transient time, the error trajectory ${e}(t)$ = ${x}(t) - {x_{d}}(t)$ converges to zero in both cases as shown in Fig. \ref{fig:sim}(b) and \ref{fig:sim}(f) respectively. Note that the control input ${u} = {\dot{\theta}}$ remain within the maximum joint velocity limits shown in Fig. \ref{fig:sim}(c) and Fig. \ref{fig:sim}(g). However unlike in optimal regulation, it does not converge to zero as the velocity compensation is required for perfect tracking control. 
The time history plot of the associated cost function is shown in shown in Fig. \ref{fig:sim}(d,h).
 The experimental validation for the corresponding target pose was performed using a real UR 10 robot manipulator. The results from Fig. \ref{fig:exp} shows that the end effector of UR 10 robot manipulator successfully reaches the target pose under the proposed control scheme and successfully follows the reference time varying circular trajectory starting from a random seed pose. 
The time history plot of the associated cost function is shown in Fig. \ref{fig:exp}(d,h). 
Unlike in simulations, the control effort shows some chatter due to the inertia of the real hardware.\vspace{-2mm}

\subsection{Quantitative Test Comparison} 
In order to quantify the performance of the proposed optimal kinematic control, an automated test process was used where  a large, statistically-valid number of random samples were used as inputs. Kinematic models of Universal Robot (UR) 10 was used to demonstrate the tests.
The quantitative test methodology for comparing the proposed method for kinematic control against the state-of-the-art kinematic control using an RNN $\cite{li2017kinematic}$ and Singluar Value Filtering (SVF) approach \cite{colome2012redundant} is entailed.

\quad First, a total of $1,000$ random samples and $1000$ different feasible elliptical trajectories were generated in the sample space for regulation and tracking respectively. The sample space is a cuboid volume of task space within robot's reach and every sample is a pair of seed pose and target pose in the case of regulation.
The three kinematic control schemes mentioned above are tested and compared on trajectory cost. 

The trajectory cost is defined as normalized total cost: $ V/N$ , where $ V=\int_t^\infty{e(t)^T{Q}e(t)+u(t)^TRu(t)+\dot{u}(t)^TR\dot{u}(t)}dt $ for the same $Q=I_{3\times 3}$ and $R=I_{6\times 6}$ and $N$ is the number of total sample instances.
The design parameters were selected such that the maximum control effort is the same for all three approaches.

\begin{tabular}{|c|c|c|}
 \hline
Controller&Trajectory cost & Trajectory Cost\\
&Regulation&Tracking\\
  \hline
 Controller \cite{li2017kinematic}& $49.74$& $14.39$\\
  \hline
 Controller \cite{colome2012redundant}&$48.32$ & $32.21$ \\
  \hline
 Proposed & $5.78$ & $3.30$\\
 \hline
 \end{tabular}
  \captionof{table}{Comparisons of different algorithms for Kinematic Control of UR10 Robot Manipulator}
  %$\mathbf{{\theta}}\in \mathbb{R}^{6}$ in the shown configuration} \label{tab:title} 
%   \end{minipage}
The cost matrix contain two terms. One for the measure of optimal control action and another for smoothness of the motion. The comparison from Table I shows that the proposed kinematic control has the optimal control action than the state-of-the-art kinematic control.

 \vspace{-3mm}
\section{Conclusion}
 \vspace{-2mm}
In this work, we have designed an optimal kinematic controller for a robot manipulator using SNAC framework. A simple critic weight update law was proposed which ensured that the closed loop system becomes stable in the sense of Lyapunov while following an optimal trajectory. The robot was expected to reach a target position or follow a time varying trajectory in the task space while optimizing a global cost function. Using the proposed optimal regulation and optimal tracking framework in the context of SNAC, it has been experimentally demonstrated that robot performs desired tasks with optimal cost as evident from Table 1.

\nocite{*}
\bibliographystyle{IEEEtran}
\bibliography{main}

\begin{thebibliography}{10}
\providecommand{\url}[1]{#1}
\csname url@samestyle\endcsname
\providecommand{\newblock}{\relax}
\providecommand{\bibinfo}[2]{#2}
\providecommand{\BIBentrySTDinterwordspacing}{\spaceskip=0pt\relax}
\providecommand{\BIBentryALTinterwordstretchfactor}{4}
\providecommand{\BIBentryALTinterwordspacing}{\spaceskip=\fontdimen2\font plus
\BIBentryALTinterwordstretchfactor\fontdimen3\font minus
  \fontdimen4\font\relax}
\providecommand{\BIBforeignlanguage}[2]{{%
\expandafter\ifx\csname l@#1\endcsname\relax
\typeout{** WARNING: IEEEtran.bst: No hyphenation pattern has been}%
\typeout{** loaded for the language `#1'. Using the pattern for}%
\typeout{** the default language instead.}%
\else
\language=\csname l@#1\endcsname
\fi
#2}}
\providecommand{\BIBdecl}{\relax}
\BIBdecl

\bibitem{Swagat10}
S.~Kumar, L.~Behera, and T.~McGinnity, ``Kinematic control of a redundant
  manipulator using inverse-forward adaptive scheme with a ksom based hint
  generator,'' \emph{Robotics and Autonomous Systems}, vol.~58, no.~5, pp.
  622--633, 2010.

\bibitem{Prem10}
P.~Prem~Kumar and L.~Behera, ``Visual servoing of a redundant manipulator with
  jacobian matrix estimation using self-organizing map,'' \emph{Robotics and
  Autonomous Systems}, vol.~58, no.~8, pp. 978--990, 2010.

\bibitem{li2017kinematic}
S.~Li, Y.~Zhang, and L.~Jin, ``Kinematic control of redundant manipulators
  using neural networks,'' \emph{IEEE transactions on neural networks and
  learning systems}, vol.~28, no.~10, pp. 2243--2254, 2017.

\bibitem{indrazno14}
I.~Sirazuddin, L.~Behera, T.~McGinnity, and S.~Coleman, ``Image based visual
  servoing of a 7 dof robot manipulator using an adaptive distributed fuzzy pd
  controller,'' \emph{IEEE/ASME Trans on Mechatronics}, vol.~19, no.~2, pp.
  512--523, 2014.

\bibitem{lewis2012optimal}
F.~L. Lewis, D.~Vrabie, and V.~L. Syrmos, \emph{Optimal control}.\hskip 1em
  plus 0.5em minus 0.4em\relax John Wiley \& Sons, 2012.

\bibitem{keerthi1985existence}
S.~Keerthi and E.~Gilbert, ``An existence theorem for discrete-time
  infinite-horizon optimal control problems,'' \emph{IEEE Transactions on
  Automatic Control}, vol.~30, no.~9, pp. 907--909, 1985.

\bibitem{bertsekas2005dynamic}
D.~P. Bertsekas, D.~P. Bertsekas, D.~P. Bertsekas, and D.~P. Bertsekas,
  \emph{Dynamic programming and optimal control}.\hskip 1em plus 0.5em minus
  0.4em\relax Athena scientific Belmont, MA, 2005, vol.~1, no.~3.

\bibitem{chen2008generalized}
Z.~Chen and S.~Jagannathan, ``Generalized hamilton--jacobi--bellman
  formulation-based neural network control of affine nonlinear discrete-time
  systems,'' \emph{IEEE Transactions on Neural Networks}, vol.~19, no.~1, pp.
  90--106, 2008.

\bibitem{si2004handbook}
J.~Si, A.~Barto, W.~Powell, and D.~Wunsch, ``Handbook of learning and approx.
  dynamics prog,'' 2004.

\bibitem{padhi2006single}
R.~Padhi, N.~Unnikrishnan, X.~Wang, and S.~Balakrishnan, ``A single network
  adaptive critic (snac) architecture for optimal control synthesis for a class
  of nonlinear systems,'' \emph{Neural Networks}, vol.~19, no.~10, pp.
  1648--1660, 2006.

\bibitem{patchaikani2012single}
P.~K. Patchaikani, L.~Behera, and G.~Prasad, ``A single network adaptive
  critic-based redundancy resolution scheme for robot manipulators,''
  \emph{IEEE Transactions on Industrial Electronics}, vol.~59, no.~8, pp.
  3241--3253, 2012.

\bibitem{christoph2014}
C.~S. . T. B. . J. B. . J. P. .~H. Ulbrich, ``Predictive online inverse
  kinematics for redundant manipulators,'' in \emph{ICRA Proceedings of the
  2014 IEEE International Conference on}.\hskip 1em plus 0.5em minus
  0.4em\relax IEEE, 2014, pp. 5056--5061.

\bibitem{Ren2014}
R.~C. L. . T.-W. L. . Y.-H. Tsai, ``Analytical inverse kinematic solution for
  modularized 7-dof redundant manipulators with offsets at shoulder and
  wrist,'' in \emph{IROS Proceedings of the 2014 IEEE/RSJ International
  Conference on}.\hskip 1em plus 0.5em minus 0.4em\relax IEEE, 2014, pp.
  516--521.

\bibitem{sqp_icra17}
M.~Giftthaler, F.~Farshidian, T.~Sandy, L.~Stadelmann, and J.~Buchli,
  ``Efficient kinematic planning for mobile manipulators with non-holonomic
  constraints using optimal control,'' in \emph{2017 IEEE International
  Conference on Robotics and Automation (ICRA)}, May 2017, pp. 3411--3417.

\bibitem{hollerbach1985redundancy}
J.~Hollerbach and K.~Suh, ``Redundancy resolution of manipulators through
  torque optimization,'' in \emph{Robotics and Automation. Proceedings. 1985
  IEEE International Conference on}, vol.~2.\hskip 1em plus 0.5em minus
  0.4em\relax IEEE, 1985, pp. 1016--1021.

\bibitem{kim94}
S.-W. Kim, K.-B. Park, and J.-J. Lee, ``Redundancy resolution of robot
  manipulators using optimal kinematic control,'' in \emph{Robotics and
  Automation, 1994. Proceedings., 1994 IEEE International Conference on}.\hskip
  1em plus 0.5em minus 0.4em\relax IEEE, 1994, pp. 683--688.

\bibitem{hollerbach89}
D.~P. Martin, J.~Baillieul, and J.~M. Hollerbach, ``Resolution of kinematic
  redundancy using optimization techniques,'' \emph{IEEE Transactions on
  Robotics and Automation}, vol.~5, no.~4, pp. 529--533, 1989.

\bibitem{maciejewski1989kinetic}
A.~Maciejewski, ``Kinetic limitations on the use of redundancy in robotic
  manipulators,'' in \emph{Robotics and Automation, 1989. Proceedings., 1989
  IEEE International Conference on}.\hskip 1em plus 0.5em minus 0.4em\relax
  IEEE, 1989, pp. 113--118.

\bibitem{o2002divergence}
K.~A. O'Neil, ``Divergence of linear acceleration-based redundancy resolution
  schemes,'' \emph{IEEE Transactions on Robotics and Automation}, vol.~18,
  no.~4, pp. 625--631, 2002.

\bibitem{cocuzza2011novel}
S.~Cocuzza, I.~Pretto, and S.~Debei, ``Novel reaction control techniques for
  redundant space manipulators: Theory and simulated microgravity tests,''
  \emph{Acta Astronautica}, vol.~68, no. 11-12, pp. 1712--1721, 2011.

\bibitem{kiumarsi2015actor}
B.~Kiumarsi and F.~L. Lewis, ``Actor--critic-based optimal tracking for
  partially unknown nonlinear discrete-time systems,'' \emph{IEEE transactions
  on neural networks and learning systems}, vol.~26, no.~1, pp. 140--151, 2015.

\bibitem{wang2018neural}
D.~Wang, D.~Liu, Y.~Zhang, and H.~Li, ``Neural network robust tracking control
  with adaptive critic framework for uncertain nonlinear systems,''
  \emph{Neural Networks}, vol.~97, pp. 11--18, 2018.

\bibitem{dierks2010optimal}
T.~Dierks and S.~Jagannathan, ``Optimal control of affine nonlinear
  continuous-time systems,'' in \emph{American Control Conference (ACC),
  2010}.\hskip 1em plus 0.5em minus 0.4em\relax IEEE, 2010, pp. 1568--1573.

\bibitem{vamvoudakis2016control}
K.~Vamvoudakis and S.~Jagannathan, \emph{Control of Complex Systems: Theory and
  Applications}.\hskip 1em plus 0.5em minus 0.4em\relax Butterworth-Heinemann,
  2016.

\bibitem{selmic1999backlash}
R.~R. Selmic and F.~L. Lewis, ``Backlash compensation in nonlinear systems
  using dynamic inversion by neural networks,'' in \emph{Control Applications,
  1999. Proceedings of the 1999 IEEE International Conference on},
  vol.~2.\hskip 1em plus 0.5em minus 0.4em\relax IEEE, 1999, pp. 1163--1168.

\bibitem{colome2012redundant}
A.~Colom{\'e} and C.~Torras, ``Redundant inverse kinematics: Experimental
  comparative review and two enhancements,'' in \emph{Intelligent Robots and
  Systems (IROS), 2012 IEEE/RSJ International Conference on}.\hskip 1em plus
  0.5em minus 0.4em\relax IEEE, 2012, pp. 5333--5340.

\bibitem{vamvoudakis2011multi}
K.~G. Vamvoudakis and F.~L. Lewis, ``Multi-player non-zero-sum games: Online
  adaptive learning solution of coupled hamilton--jacobi equations,''
  \emph{Automatica}, vol.~47, no.~8, pp. 1556--1569, 2011.

\bibitem{dierks2009optimal}
T.~Dierks and S.~Jagannathan, ``Optimal tracking control of affine nonlinear
  discrete-time systems with unknown internal dynamics,'' in \emph{Decision and
  Control, 2009 held jointly with the 2009 28th Chinese Control Conference.
  CDC/CCC 2009. Proceedings of the 48th IEEE Conference on}.\hskip 1em plus
  0.5em minus 0.4em\relax IEEE, 2009, pp. 6750--6755.

\bibitem{vrabie2009neural}
D.~Vrabie and F.~Lewis, ``Neural network approach to continuous-time direct
  adaptive optimal control for partially unknown nonlinear systems,''
  \emph{Neural Networks}, vol.~22, no.~3, pp. 237--246, 2009.

\bibitem{slotine1991applied}
J.-J.~E. Slotine, W.~Li \emph{et~al.}, \emph{Applied nonlinear control}.\hskip
  1em plus 0.5em minus 0.4em\relax Prentice hall Englewood Cliffs, NJ, 1991,
  vol. 199, no.~1.

\end{thebibliography}

\end{document}